\documentclass[amssymb,amsmath,aps,floatfix,nofootinbib,showpacs,12pt,prd]{revtex4}

\usepackage{graphicx,latexsym,color,bm}
\usepackage{threeparttable}

\newcommand{\be}{\begin{equation}}
\newcommand{\ee}{\end{equation}}
\newcommand{\bea}{\begin{eqnarray}}
\newcommand{\eea}{\end{eqnarray}}

\begin{document}

\title{CosRayMC: a global fitting method in studying the properties of the new sources of cosmic e$^{\pm}$ excesses}

\author{Jie Liu$^a$}
\author{Qiang Yuan$^b$}
\author{Xiao-Jun Bi$^{b}$}
\author{Hong Li$^{a,c}$}
\author{Xinmin Zhang$^{a,c}$}

\affiliation{$^a$Theoretical Physics Division, Institute of High
Energy Physics, Chinese Academy of Science, P.O.Box 918-4, Beijing
100049, P.R.China}

\affiliation{$^b$Key Laboratory of Particle Astrophysics,
Institute of High Energy Physics, Chinese Academy of Science,
P.O.Box 918-3, Beijing 100049, P.R.China}

\affiliation{$^c$Theoretical Physics Center for Science Facilities
(TPCSF), Chinese Academy of Science, Beijing 100049, P.R.China}

%\date{\today}

\begin{abstract}

Recently PAMELA collaboration published the cosmic nuclei and electron 
spectra with high precision, together with the cosmic antiproton data 
updated, and the Fermi-LAT collaboration also updated the measurement of the 
total $e^+e^-$ spectrum to lower energies. In this paper we develop a 
Markov Chain Monte Carlo (MCMC) package {\it CosRayMC}, based on the 
GALPROP cosmic ray propagation model to study the implications of these 
new data. It is found that if only the background electrons and 
secondary positrons are considered, the fit is very bad with 
$\chi_{\rm red}^2 \approx 3.68$. Taking into account the extra $e^+e^-$
sources of pulsars or dark matter annihilation we can give much better 
fit to these data, with the minimum $\chi_{\rm red}^2 \approx 0.83$. 
This means the extra sources are necessary with a very high significance 
in order to fit the data. However, the data show little difference 
between pulsar and dark matter scenarios. Both the background and extra 
source parameters are well constrained with this MCMC method. Including 
the antiproton data, we further constrain the branching ratio of dark 
matter annihilation into quarks $B_q<0.5\%$ at $2\sigma$ confidence
level. The possible systematical uncertainties of the present
study are discussed.
%With more high quality data and better understanding of
%the inputs, it is the best way to extract useful information from
%the data with the global fitting method.

\end{abstract}

%95.35.+d: Dark matter
%96.50.S-: Cosmic rays

\maketitle

\section{Introduction}

A very interesting progress made in the recent years in cosmic ray
(CR) physics is the discovery of the excesses of positrons and
electrons by several space- and ground-based experiments
\cite{2009Natur.458..607A,2008Natur.456..362C,
2008PhRvL.101z1104A,2009A&A...508..561A,2009PhRvL.102r1101A}. The
positron and electron excesses challenge the traditional understanding 
of CR background. Many theoretical models were proposed to explain 
these new phenomena, including the astrophysical scenarios (e.g.,
\cite{2009PhRvL.103e1101Y,2009JCAP...01..025H,2009A&A...501..821D,
2009PhRvL.103k1302S,2009ApJ...700L.170H,2009PhRvL.103e1104B}, and see 
\cite{2010IJMPD..19.2011F} for a review) and the dark matter (DM) scenario
(e.g., \cite{2008PhRvD..78j3520B,2009PhLB..672..141B,2009NuPhB.813....1C,
2009PhRvD..79b3512Y})
%for review see \cite{2009MPLA...24.2139H}).

Most recently Fermi-LAT team reported the updated measurements of
the total spectrum of electrons and positrons, extending to energies
as low as several GeV \cite{2010PhRvD..82i2004A}. PAMELA collaboration
updated the observation of the $\bar{p}/p$ ratio and the absolute
antiproton flux \cite{2010PhRvL.105l1101A}, and reported the measurement
of the pure electron spectrum at the first time \cite{2011PhRvL.106t1101A}.
The antiproton data extend to $180$ GeV, without any hint of deviation
from the background contribution \cite{2010PhRvL.105l1101A}. For the
PAMELA electron data, although there is no significant spectral feature
above $30$ GeV other than a single power-law, it is also consistent
with models including extra $e^+e^-$ sources to explain the
positron excess \cite{2011PhRvL.106t1101A}. There is also hint of
hardening of the electron spectrum compared with the low energy part
($<20$ GeV), although the solar modulation may be important at these
low energies.

Since the accumulation of high-quality CR data, it is now important 
to extract more information from these data, i.e., estimating the CR 
background and the possible extra source parameters. Previously, one
always constrained one or two parameters with other parameters fixed.
This may lead to biased results, especially, when the parameters 
are strongly correlated. The global fitting procedure searches the maximum 
likelihood in the multiple dimensional parameter space rather than 
a reduced one and can give the posterior distribution by marginalizing
other parameters in the Bayesian approach. The Markov Chain Monte Carlo 
(MCMC) procedure, whose computational time scales approximately 
linearly with the number of parameters, makes it possible to survey in 
a very large parameter space with the least computational cost.

In our previous studies \cite{2010PhRvD..81b3516L,2009arXiv0911.1002L},
we employed the MCMC method to fit the parameters of the DM scenario
as well as the background parameters proposed to explain the $e^+e^-$
excesses. However, the propagation of CRs is treated with a semi-analytical
way following Refs. \cite{2007A&A...462..827L,2008A&A...479..427L,
2006astro.ph..9522M}. A more precise description of the CR propagation
is given by the numerical models, such as GALPROP \cite{1998ApJ...509..212S}
and DRAGON \cite{2008JCAP...10..018E}, in which most of the relevant
physical processes are taken into account, and the realistic astrophysical
inputs like the interstellar medium (ISM) and the interstellar radiation
field (ISRF) are adopted. There are many parameters in the CR propagation
model, and it is very difficult to have a full and systematical survey
of the parameter space. After embedding the numerical CR propagation
tool into the MCMC sampler, we can use it to constrain the model parameters
in a more efficient way. Recently there are also several other works
using the MCMC method to study the CR propagation \cite{2009A&A...497..991P,
2010A&A...516A..66P,2011ApJ...729..106T}.

In this work, we develop a package {\it CosRayMC} (Cosmic Ray MCMC),
which is comprised of GALPROP, PYTHIA \cite{2006JHEP...05..026S} and
MCMC sampler, to revisit the models to explain the positron and
electron excesses and derive the constraints on the model parameters
with the latest data. Two kinds of the extra $e^+e^-$ sources are
considered: the pulsar scenario and the DM annihilation scenario.
In both scenarios we consider the continuous distribution of the sources,
although it is possible that one or several nearby pulsars or DM subhalos
may explain the excesses \cite{2009PhRvL.103e1101Y,2009JCAP...01..025H,
2008arXiv0812.4457P,2009PhRvD..79j3513H,2009PhRvD..79l3517K}.
As pointed out in \cite{2009PhRvD..80f3005M}, it was unlikely that the
flux from any single pulsar was significantly larger than that from others
given a large number of known nearby and energetic pulsars. For DM subhalos
the location and mass are uncertain, which also makes the constraints of
the model parameters difficult.

This paper is organized as follows. We briefly introduce the propagation
of Galactic CRs in Sec. \ref{crprop}. The description of the methodology
is given in Sec. \ref{cosraymc}. The fitting results of the pulsar and
DM scenarios are presented in Sec. \ref{result}. Finally we give the
conclusion and discussion in Sec. \ref{summary}.

\section{Propagation of Galactic cosmic rays}
\label{crprop}

The charged particles propagate diffusively in the Galaxy due to the
scattering with random magnetic field. There are interactions between
the CR particles and the ISM and/or the ISRF, which will lead to
fragmentation, catastrophic or continuous energy losses of the particles.
For unstable nuclei the radioactive decay also needs to be taken into account.
In addition, the overall convection driven by the stellar wind and
reacceleration due to the interstellar shock will also affect the
distribution function of CRs. For each species of particles we have
a partial differential equation to describe the propagation process,
with the general form \cite{2007ARNPS..57..285S}
\begin{equation}
\frac{\partial \psi}{\partial
t} = Q({\bf x},p)+\nabla\cdot(D_{xx}\nabla \psi-{\bf
V_c}\psi)+\frac{\partial}{\partial p}p^2D_{pp}\frac{\partial}
{\partial p}\frac{1}{p^2}\psi - \frac{\partial}{\partial p}
\left[\dot{p}\psi-\frac{p}{3}(\nabla\cdot{\bf V_c}\psi)\right]-
\frac{\psi}{\tau_f}-\frac{\psi}{\tau_r}, \label{prop}
\end{equation}
where $\psi$ is the density of cosmic ray particles per unit momentum
interval, $Q({\bf x},p)$ is the source term, $D_{xx}$ is the spatial
diffusion coefficient, ${\bf V_c}$ is the convection velocity,
$D_{pp}$ is the diffusion coefficient in momentum space used to
describe the reacceleration process, $\dot{p}\equiv{\rm d}p/{\rm d}t$
is the momentum loss rate, $\tau_f$ and $\tau_r$ are time scales for
fragmentation and radioactive decay respectively. Solving the partially
coupled equations for all kinds of particles, we can get the propagated
results of the CR spectra and spatial distributions. For more details
about the terms in Eq. (\ref{prop}) please refer to the recent review
paper \cite{2007ARNPS..57..285S}.

The secondary-to-primary ratios such as B/C and (Sc+Ti+V)/Fe, and the
unstable-to-stable ratios of secondary particles such as
$^{10}$Be/$^9$Be and $^{26}$Al/$^{27}$Al, are often used to constrain
the propagation parameters because the ratios can effectively avoid
the source parameters. Then one can use the spectra of the primary
particles to derive the source parameters. There are some studies
to constrain the propagation parameters based on the currently
available data (e.g., \cite{2001ApJ...555..585M,1998ApJ...509..212S,
2010JCAP...06..022P}). In \cite{2009A&A...497..991P,2010A&A...516A..66P,
2011ApJ...729..106T} the MCMC method was adopted to fit both the
propagation and source parameters of CRs. However, due to the quality
of the observational data, the constraints on the propagation parameters
are not very effective, and there may be also large systematic uncertainties.

Since we focus on the electron and positron data in this work, which 
are not very effective to constrain the propagation parameters, we adopt the
best fitting results of the propagation parameters given in 
\cite{2011ApJ...729..106T}. The main propagation parameters are: 
$D_0=6.59\times10^{28}$ cm$^2$s$^{-1}$, $\delta=0.30$, $v_A=39.2$ 
km s$^{-1}$, $z_h=3.9$ kpc. The injection spectra of nuclei are adopted 
as $\gamma_1^n=1.91$ (below 10 GV) and $\gamma_2^n=2.40$ (above 10 GV).
What we should keep in mind is that there might be systematic errors of 
the determination of the propagation parameters (see e.g., 
\cite{2010A&A...516A..67M}), and the quantitative results of this work 
might be affected.

\begin{figure}[!htb]
\begin{center}
\includegraphics[width=0.45\columnwidth]{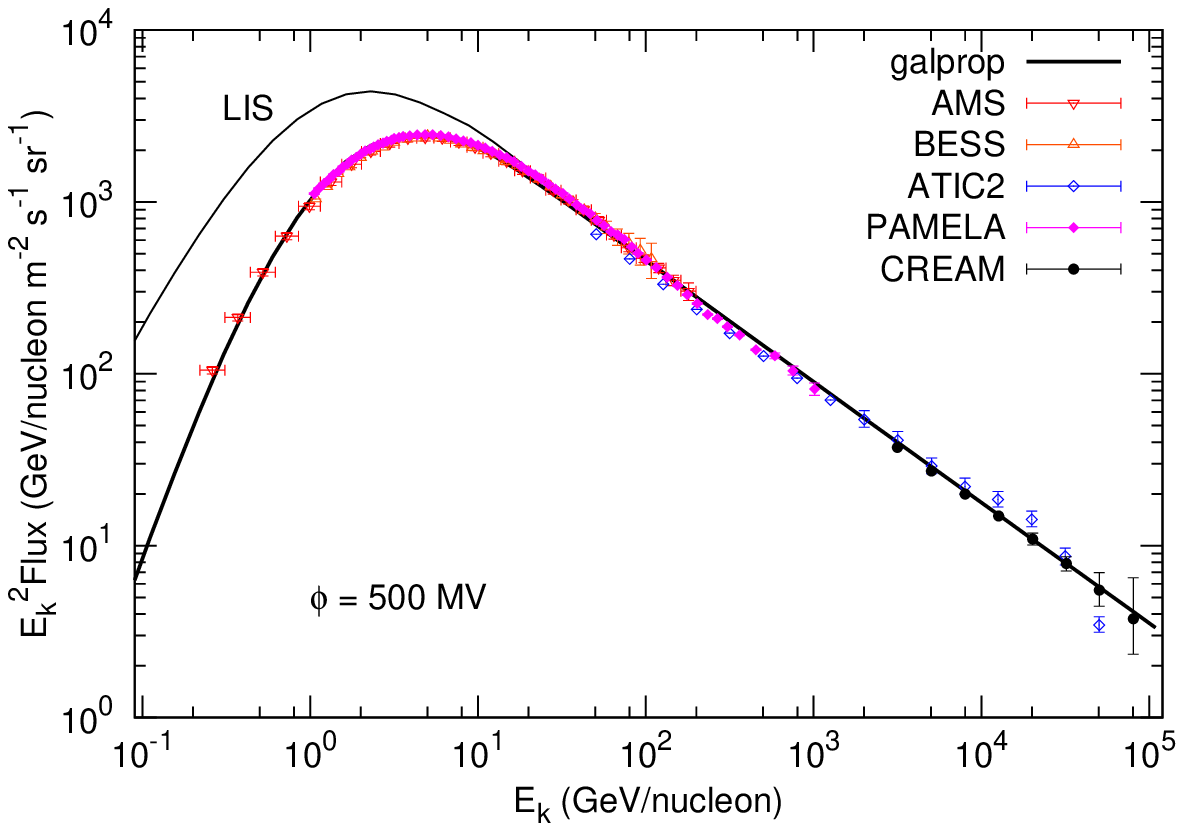}
\includegraphics[width=0.45\columnwidth]{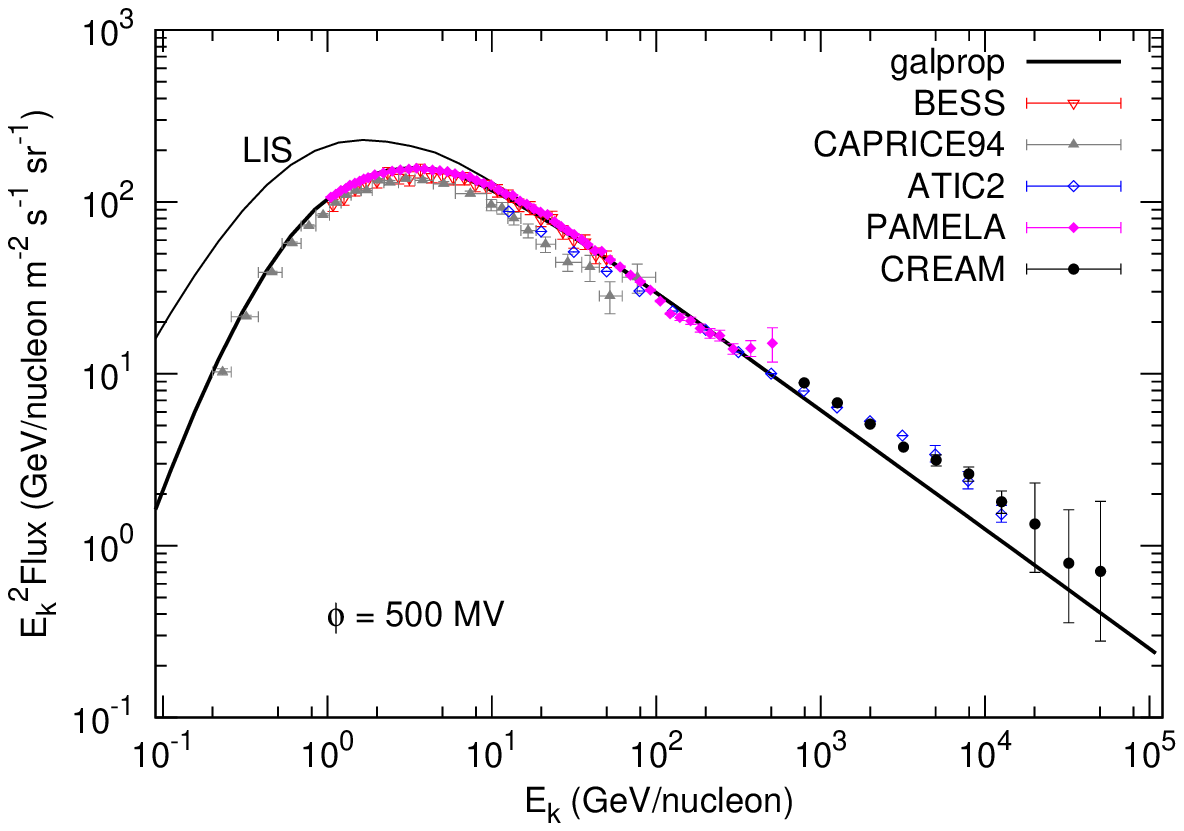}
\caption{Proton (left) and Helium (right) spectra of the GALPROP model
calculation, compared with the observational data. In each panel the
higher line labelled ``LIS'' represents the local interstellar flux,
and the lower one is the flux after the solar modulation.
References of the data are: proton --- AMS \cite{2000PhLB..490...27A},
BESS \cite{2000ApJ...545.1135S}, ATIC2 \cite{2007BRASP..71..494P},
PAMELA \cite{2011Sci...332...69A}, CREAM \cite{2010ApJ...714L..89A};
Helium --- BESS \cite{2000ApJ...545.1135S},
CAPRICE94 \cite{1999ApJ...518..457B}, ATIC2 \cite{2007BRASP..71..494P},
PAMELA \cite{2011Sci...332...69A}, CREAM \cite{2010ApJ...714L..89A}.}
\label{fig:pHe}
\end{center}
\end{figure}

We note that the newest measurements of the CR proton and Helium spectra
by CREAM \cite{2010ApJ...714L..89A} and PAMELA
\cite{2011Sci...332...69A} showed remarkable deviation of single
power-law spectra. Furthermore the spectral indices of proton and
Helium are different. Such new features challenge the traditional
understanding of the CR origin, acceleration and propagation (see e.g.
\cite{2011ApJ...729L..13O,2011arXiv1104.3357Y} for some explanations).
In the present work we keep the study in the traditional frame and do not
try to reproduce the detailed structures of the data. We find that the
above adopted injection parameters with proper adjusting
of the normalization can basically reproduce the PAMELA-CREAM proton data,
as shown in the left panel of Fig. \ref{fig:pHe}. However, when comparing
with the Helium data, the model expectation seems to be systematically
lower. We increase the relative abundance of Helium in GALPROP by $30\%$,
and give roughly consistent result with PAMELA data (right panel of Fig.
\ref{fig:pHe}). The CREAM data can still not be reproduced. Since the
contribution to the secondary particles from CR Helium is only a small
fraction, we do not think a higher Helium flux within a factor of 2 will
substantially affect our following study.

\section{Methodology}
\label{cosraymc}

\subsection{CosRayMC}

The {\it CosRayMC} (Cosmic Ray MCMC) code is built up by embedding the
CR propagation code GALPROP (GALPROP v51) into the MCMC sampling scheme. The MCMC
technique is widely applied to give multi-dimensional parameter
constraints from observational data. Following Bayes' Theorem, the
posterior probability of a model (which we refer to a series of parameters
$\vec{\theta}$) given the data (which are described by $D$) is
\begin{eqnarray}\label{eq:bayes}
{\cal P}(\vec{\theta}|D)~\propto {\cal L}(\vec{\theta}){\cal P}(\vec{\theta}),
\end{eqnarray}
where ${\mathcal L}(\vec{\theta}) = P(D|\vec{\theta})$ is the likelihood
function of the model $\vec{\theta}$ for the data $D$, and
${\cal P(\vec{\theta})}$ is the prior probability of the model parameters.
In performing {\it CosRayMC}, MCMC is employed to generate a random sample
from the posterior distribution ${\cal P}(\vec{\theta}|D)$ which are fair
samples of the likelihood surface. Based on the sample, we can get the
estimate of the mean values, the variance as well as the confidence level
of the model parameters. For details please refer to
\cite{2002PhRvD..66j3511L}.

The key improvement, compared with our previous study, is that the
calculation of the likelihood ${\cal L}(\vec{\theta})$ are given by 
calling GALPROP to simulate the mock observations. By doing so, we can 
have a more precise description of the CR propagation. And
as the better data-set is provided in the future, our code can
also be used to make a determination of the CR propagation
parameters.

For the part with DM contribution to the CRs, the PYTHIA (PYTHIA v6.4)
simulation code is employed to calculate the final spectra of
electrons, positrons and antiprotons \cite{2006JHEP...05..026S}.
Such spectra are then injected into the Galaxy and propagated with
GALPROP. The PYTHIA code is also embedded in the {\it CosRayMC}
code.

\subsection{Parameters}

We assume the injection spectrum of the background electrons to be a broken
power-law function with spectral indices $\gamma_1$/$\gamma_2$ below/above
$E_{\rm br}$. Note that for the shock acceleration scenario, the injection
spectrum of particles can not be too hard \cite{2001RPPh...64..429M}.
We set the priors that $\gamma_1>1.5$ and $\gamma_2>1.5$ in the MCMC
scanning. The normalization of background electrons $A_{\rm bkg}$,
taken as the flux of electrons at $25$ GeV, is also regarded as a free
parameter. For the background positrons and antiprotons, we adopt
the GALPROP model predicted results with the best fitting source and
propagation parameters given in \cite{2011ApJ...729..106T}. Considering
the fact that there are uncertainties about the ISM density distribution,
the hadronic interaction model, and the propagation parameters determined
from the secondary-to-primary ratio data, we will further employ
factors $c_{e^+}$ and $c_{\bar{p}}$ to rescale the absolute fluxes of
these secondary particle.

For energies below $\sim 30$ GeV the solar modulation effect is
important and needs to be considered. In this work we adopt the
force-field approximation to calculate the solar modulation
\cite{1968ApJ...154.1011G}. The modulation potential depends on
the solar activity. For the period which PAMELA works the
modulation potential is estimated to be $\phi=450-550$ MV
\cite{2011Sci...332...69A}. In our MCMC fit, the modulation
potential $\phi$ is also taken as a free parameter. Thus, for the
background model we have $7$ parameters in total
\begin{equation}
{\mathcal P}_{\rm bkg} = \{\gamma_1, \gamma_2, E_{\rm br}, A_{\rm bkg},
\phi, c_{e^+}, c_{\bar{p}}\}.
\end{equation}

Pulsars are thought to be the most natural candidates to generate high
energy positrons and electrons through the cascade of electrons accelerated
in the magnetosphere \cite{2001A&A...368.1063Z,2008arXiv0812.4457P}.
The spectrum of $e^+e^-$ escaped from the pulsars can be parameterized
as power-law with a cutoff at $E_c$, ${\rm d}N/{\rm d}E \propto
A_{\rm psr} E^{-\alpha}\exp(-E/E_c)$, where the power-law index $\alpha$
ranges from 1 to 2.2 according to the radio and gamma-ray observations
\cite{2008arXiv0812.4457P}. The cutoff energy of the injected $e^+e^-$
ranges from several tens GeV to higher than TeV, depending on the models
and parameters of the pulsars \cite{2001A&A...368.1063Z,2009PhRvD..80f3005M}.
For the spatial distribution of pulsars, we adopt the following form
\cite{2004IAUS..218..105L}
\begin{equation}
f(R,z)\propto\left(\frac{R}{R_{\odot}}\right)^a\exp\left[-\frac{b(R-
R_{\odot})}{R_{\odot}}\right]\exp\left(-\frac{|z|}{z_s}\right),
\end{equation}
where $R_{\odot}=8.5$ kpc is the distance of solar system from the
Galactic center, $z_s\approx 0.2$ kpc is the scale height of the pulsar
distribution, $a=2.35$ and $b=5.56$. There will be a further normalization
factor $A_{\rm psr}$.

Alternatively, DM annihilation or decay models are widely employed to 
explain the $e^+e^-$ excesses. Since to date one can not distinguish DM 
annihilation from decay with the $e^+e^-$ data \cite{2010PhRvD..81b3516L, 
2009arXiv0911.1002L}, we only consider the annihilation scenario here. 
Considering the non-excess of PAMELA antiproton data, the DM annihilation 
final states need to be lepton-dominated \cite{2009NuPhB.813....1C,
2009PhRvD..79b3512Y}. Similar as the previous works
\cite{2010PhRvD..81b3516L,2009arXiv0911.1002L}, we do not employ the 
annihilation final states based on any DM models, but to constrain the 
final states from the data we use a model-independent way instead. Such 
constraints will be helpful for the understanding of the DM particle nature.

The annihilation final states are assumed to be two-body $e^+e^-$,
$\mu^+\mu^-$, $\tau^+\tau^-$ and $q\bar{q}$\footnote{Since the antiproton
productions of various quark flavors do not differ much from each other,
here we do not distinguish quark flavors but adopt $u\bar{u}$ channel
as a typical one of quark final states.}, with branching ratios
$B_e$, $B_{\mu}$, $B_{\tau}$ and $B_q$ respectively. It was also
shown that the interactions with low-mass intermediate bosons which
then decay to lepton pairs could fit the CR $e^+e^-$ data well
\cite{2009PhRvL.103c1103B,2009PhRvD..80l3518C}. However, since the
spectral shapes of the electrons and positrons from these models do
not have distinct properties which can be easily distinguished from
$\mu^+\mu^-$ and $\tau^+\tau^-$, we do not include such final
states in this study. The resulting positron, electron and antiproton
spectra of these final states are calculated using the PYTHIA
simulation package \cite{2006JHEP...05..026S}.

The DM density profile is assumed to be an Einasto type \cite{Einasto:1965}
\begin{equation}
\rho(r)=\rho_{-2}\exp\left[-\frac{2}{\alpha}\left(\frac{r^{\alpha}}
{r^{\alpha}_{-2}}-1\right)\right],
\end{equation}
where $\alpha\approx 0.17$, $r_{-2}\approx15.6$ kpc and $\rho_{-2}\approx 
0.14$ GeV cm$^{-3}$ according to the recent high resolution simulation 
{\it Aquarius} \cite{2010MNRAS.402...21N}. The corresponding local density 
of DM is about $0.44$ GeV cm$^{-3}$, which is consistent with the results 
of a larger local density derived in recent studies \cite{2010JCAP...08..004C,
2010A&A...523A..83S,2010PhRvD..82b3531P}. Then the source function
of positron produced from DM annihilation is
\begin{equation}
q(E,r)=\frac{\langle\sigma v\rangle}{2m_{\chi}^2}\sum_{f}B_f
\left.\frac{{\rm d}N}{{\rm d}E}\right|_f\times \rho^2(r),
\end{equation}
where $m_{\chi}$ is the mass of the DM particle, $\langle\sigma v\rangle$
is the velocity weighted annihilation cross section, $\left.\frac{{\rm d}N}
{{\rm d}E}\right|_f$ is the positron production rate from channel $f$ of
one annihilation.

The most general parameter space is summerized as below.
\begin{eqnarray}
\mathcal{P}_{tol}=\begin{cases}\{\mathcal{P}_{\rm bkg}\},&\mbox{background} \\
              \{\mathcal{P}_{\rm bkg},A_{\rm psr},\alpha,E_c\},&\mbox{pulsar}\\
              \{\mathcal{P}_{\rm bkg},c_{\bar p},m_{\chi},\langle\sigma
              v\rangle,B_e,B_{\mu},B_{\tau},B_u\}, &\mbox{DM}.
\end{cases}
\end{eqnarray}
For the DM annihilation scenario, we further consider two cases. One is
to use the $e^+e^-$ data only in the fit, which can be directly compared
with the pulsar scenario. In the other case, the antiproton data are
included in order to constrain the coupling to quarks of DM. In the case
without the antiproton data, $B_u$ and $c_{\bar p}$ are set to zero.

\section{Results}
\label{result}

\subsection{Background}

First we run a fit for the pure background contribution. The data included
in the fit are PAMELA positron fraction \cite{2009Natur.458..607A},
PAMELA electron \cite{2011PhRvL.106t1101A}, Fermi-LAT total $e^+e^-$
\cite{2010PhRvD..82i2004A}, and HESS total $e^+e^-$
\cite{2008PhRvL.101z1104A,2009A&A...508..561A}.

For the PAMELA positron fraction data, we only employ the data points
with $E>5$ GeV in the fit. For the lower energy part we can see that
the PAMELA data are inconsistent with the previous measurements. This
may be due to more complicated solar modulation effect or even the
charge dependent modulation \cite{1996ApJ...464..507C,2009NJPh...11j5021B,
2010arXiv1011.4843B}. It was shown that either a model with different
modulation potentials for positive and negative charged particle
\cite{2009NJPh...11j5021B} or a detailed Monte Carlo approach to solve
the stochastic differential equations of the $e^+e^-$ motion
\cite{2010arXiv1011.4843B} can give consistent description to the PAMELA
and previous data. The demodulated interstellar spectra of electrons and
positrons are also consistent with the conventional CR background model
expectation. Here we do not consider the detailed solar modulation models.
But note that the solar modulation model may affect the quantitative
fitting results. This is one kind of systematical errors.

\begin{figure}[!htb]
\begin{center}
\includegraphics[width=0.45\columnwidth]{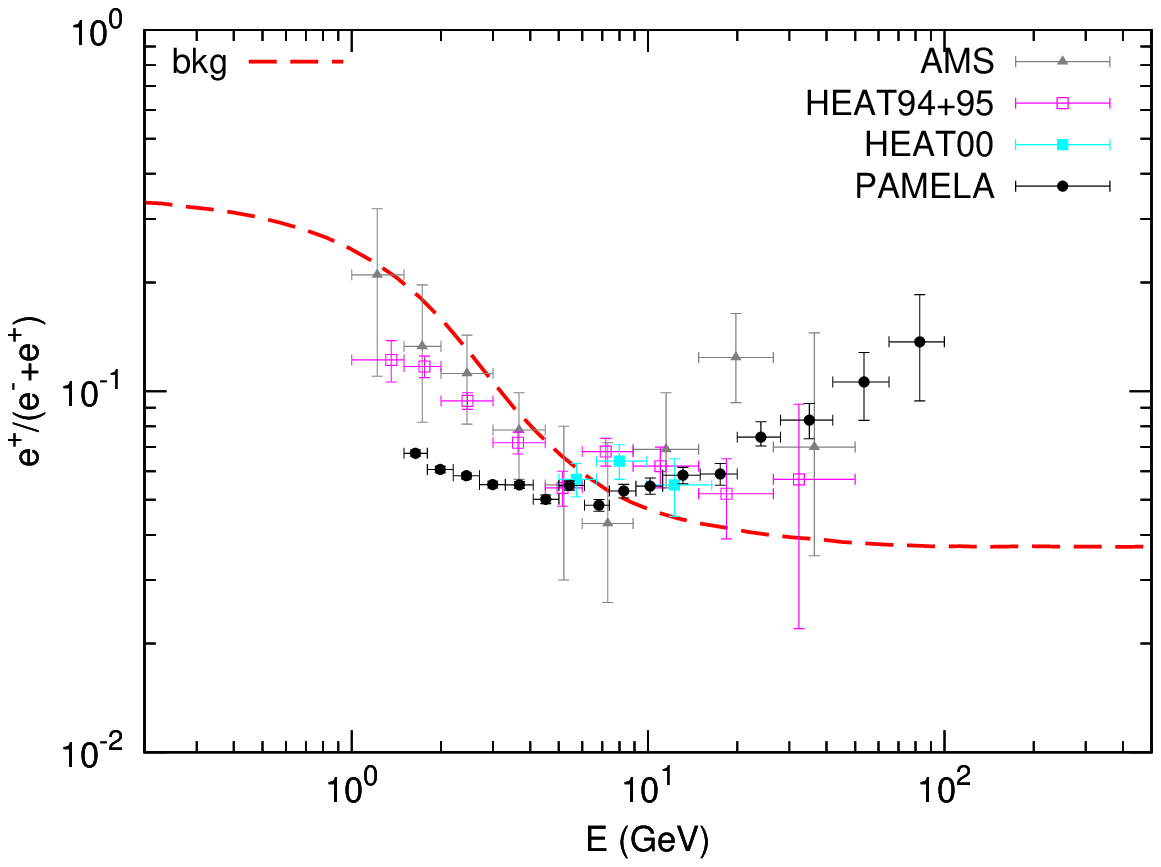}
\includegraphics[width=0.45\columnwidth]{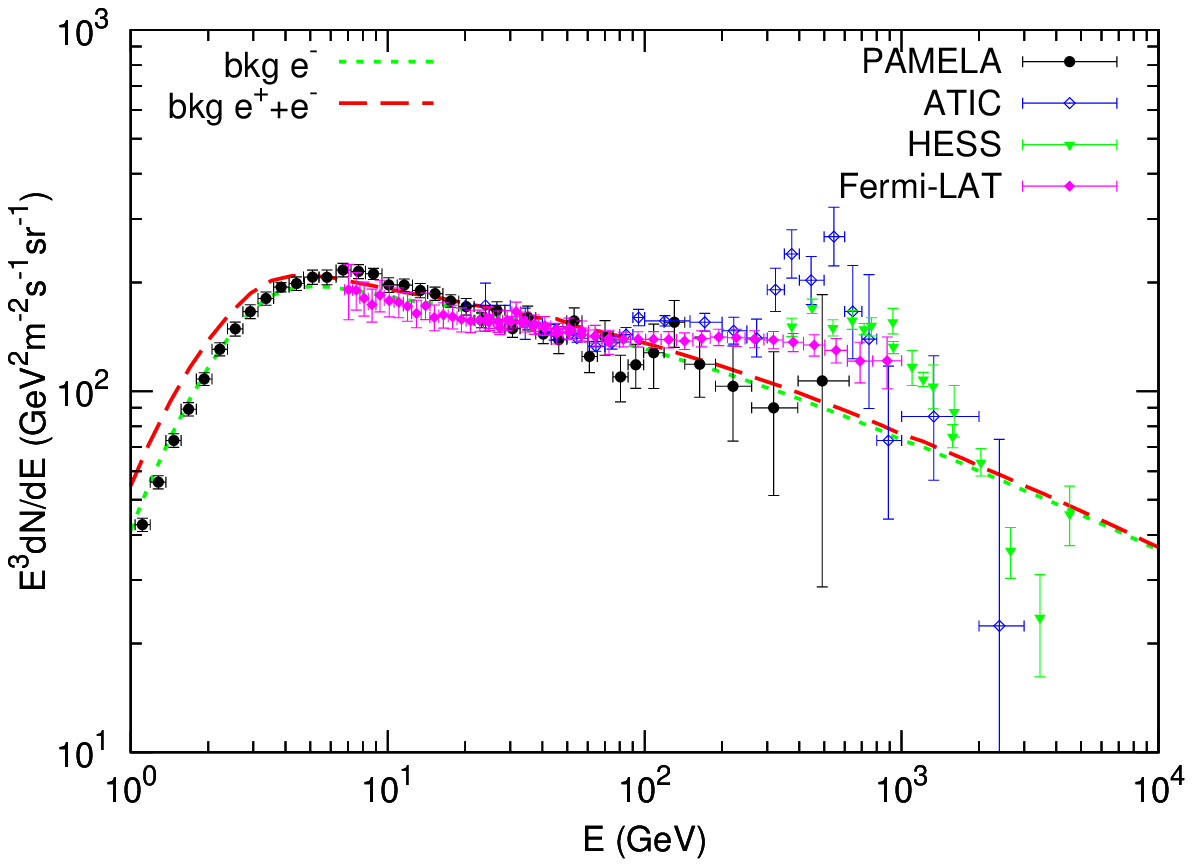}
\caption{Best-fit positron fraction (left) and electron spectra (right)
of the pure background case. Note that in the right panel the PAMELA data
are for pure electrons, while other data are for the sum of electrons and
positrons. References of the observational data are: positron fraction ---
AMS \cite{2007PhLB..646..145A}, HEAT94+95 \cite{1997ApJ...482L.191B},
HEAT00 \cite{2001ICRC....5.1687C}, PAMELA \cite{2009Natur.458..607A};
electron --- PAMELA \cite{2011PhRvL.106t1101A},
ATIC \cite{2008Natur.456..362C}, HESS \cite{2008PhRvL.101z1104A,
2009A&A...508..561A}, Fermi-LAT \cite{2010PhRvD..82i2004A}.
\label{fig:epm_bkg}}
\end{center}
\end{figure}

The best-fit results of the positron fraction and electron (or $e^+e^-$)
spectra are shown in Fig. \ref{fig:epm_bkg}. The best-fit parameters are
compiled in Table \ref{table:par}, with the minimum $\chi^2_{\rm red}
\approx 415.4/113=3.68$. Such a large reduced $\chi^2$ means that the
fit is far from acceptable but systematics dominated. Using a $\chi^2$
goodness-of-fit test we find that the background only case is rejected
with a very high significance $\sim 20\sigma$. That is to say the data
strongly favor the existence of additional degrees of freedom. From
Fig. \ref{fig:epm_bkg} we can clearly see that the background model
under-estimates both the positron fraction and the total $e^+e^-$
fluxes. The extra sources of $e^+e^-$ are necessary to explain
simultaneously the positron fraction and total $e^+e^-$ data.

\begin{table}
\begin{threeparttable}
\centering
\caption{Fitting parameters with $1\sigma$ uncertainties or $2\sigma$
limits. Note that for the ``bkg'' case the reduced $\chi^2$ is too large
that the uncertainties of the parameters should not be statistically
meaningful.}
\begin{tabular}{ccccc}
\hline
\hline
& bkg & bkg+pulsar & bkg+DM (without $\bar p$) & bkg+DM (with $\bar p$) \\
\hline
$\gamma_1$ & $<1.535 (95\% C.L.)$ & $1.630_{-0.083}^{+0.080}$ &
$1.614_{-0.070}^{+0.072}$ & $<1.610 (95\% C.L.)$\\
\hline
$\gamma_2$ & $2.559\pm0.008$ & $2.738\pm0.021$ & $2.730\pm0.018$ &
$2.706\pm0.013$\\
\hline
$\log(A_{\rm bkg})$~\tnote{1} & $-8.958\pm0.004$ & $-8.981\pm0.010$ & $-8.984\pm0.008$
& $-8.997\pm0.006$\\
\hline
$E_{\rm br} (GeV)$ & $3.554_{-0.129}^{+0.133}$ & $3.969_{-0.291}^{+0.319}$ &
$3.991_{-0.264}^{+0.290}$ & $ 4.283_{-0.259}^{+0.246}$\\
\hline
$\phi (GV)$ & $0.333\pm0.019$ & $0.499\pm0.071$ & $0.477_{-0.061}^{+0.063}$ &
$0.371\pm0.037$\\
\hline
$c_{e^+}$ & $1.473\pm0.037$ & $1.596\pm0.111$ &
$1.516_{-0.080}^{+0.079}$ & $1.394\pm0.053$ \\
\hline
$c_{\bar p}$ & --- & --- & --- & $1.210\pm0.045$ \\
\hline
$\log(A_{\rm psr})$~\tnote{2} & --- & $-28.36_{-0.65}^{+0.63}$ & --- & ---\\
\hline
$\alpha$ & --- & $1.201_{-0.129}^{+0.120}$ & --- & --- \\
\hline
$E_c$ & --- & $0.797_{-0.164}^{+0.154}$ & --- & --- \\
\hline
$M_\chi (TeV)$ & --- & --- & $2.370_{-0.385}^{+0.465}$ &
$2.341_{-0.391}^{+0.492}$ \\
\hline
$\log[\sigma v (cm^3 s^{-1})]$ & --- & --- & $-22.33\pm0.13$ & $-22.34\pm0.13$\\
\hline
$B_e$ & --- & --- & $<0.369(95\% C.L.)$ & $<0.379(95\%C.L.)$\\
\hline
$B_\mu$ & --- & --- & $<0.381(95\%C.L.)$ & $<0.334(95\%C.L.)$\\
\hline
$B_\tau$ & --- & --- &  $0.688_{-0.147}^{+0.142}$ &
$0.713_{-0.152}^{+0.141}$\\
\hline
$B_u$ & --- & --- & --- & $<0.005(95\%C.L.)$\\
\hline
$\chi^2/d.o.f$ &  $3.676$ & $0.833$ &  $0.827$ & $1.078$\\
\hline
\end{tabular}
\begin{tablenotes}
\footnotesize
\item[1] $A_{\rm bkg}$ is in unit of $cm^{-2} sr^{-1} s^{-1} MeV^{-1}$
\item[2] $A_{\rm psr}$ is in unit of $cm^{-2} sr^{-1} s^{-1} MeV^{-1}$
\end{tablenotes}
\label{table:par}
\end{threeparttable}
\end{table}

The poor fit makes nonsense to discuss the physical implication of
the parameters. Therefore we leave the discussion of the parameters
in the next subsections, when the extra sources of $e^+e^-$ are
taken into account.

\subsection{Pulsar scenario}

%\begin{figure}[!htb]
%\begin{center}
%\includegraphics[width=0.9\columnwidth]{pulsar1d.ps}
%\caption{One dimensional probability distributions of the model parameters
%for the pulsar scenario.
%\label{fig:pulsar1d}}
%\end{center}
%\end{figure}

\begin{figure}[!htb]
\begin{center}
\includegraphics[width=0.9\columnwidth]{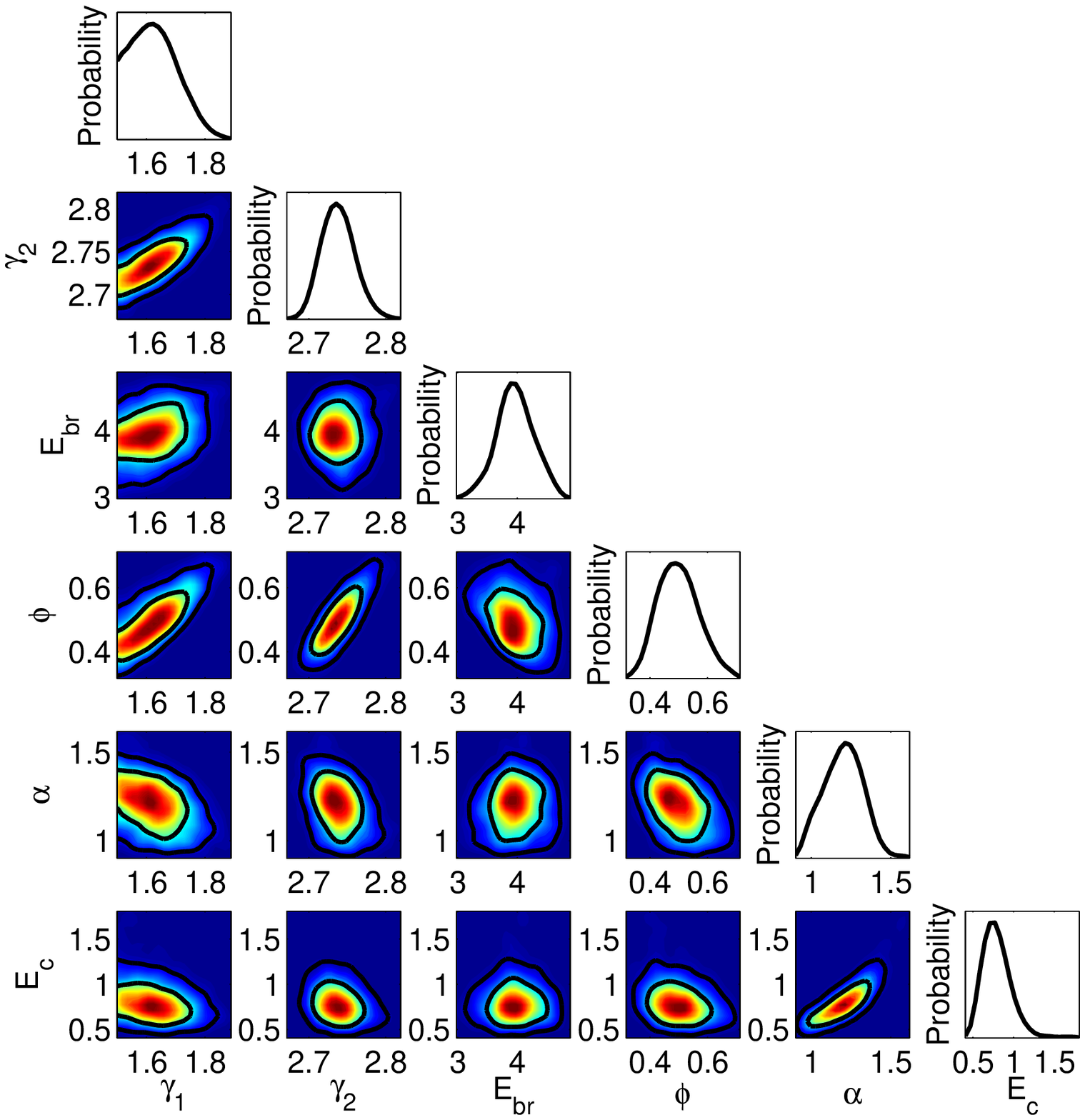}
\caption{Two dimensional constraints of part of the model
parameters for the pulsar scenario. In off-diagonal figures,
the inner contour denotes the $68\%$ confidence level (C.L.) 
while the outer contour stands for $95\%$ C.L. The diagonal figures
are one dimensional probability distributions for corresponding parameters.
\label{fig:pulsar2d}}
\end{center}
\end{figure}

\begin{figure}[!htb]
\begin{center}
\includegraphics[width=0.45\columnwidth]{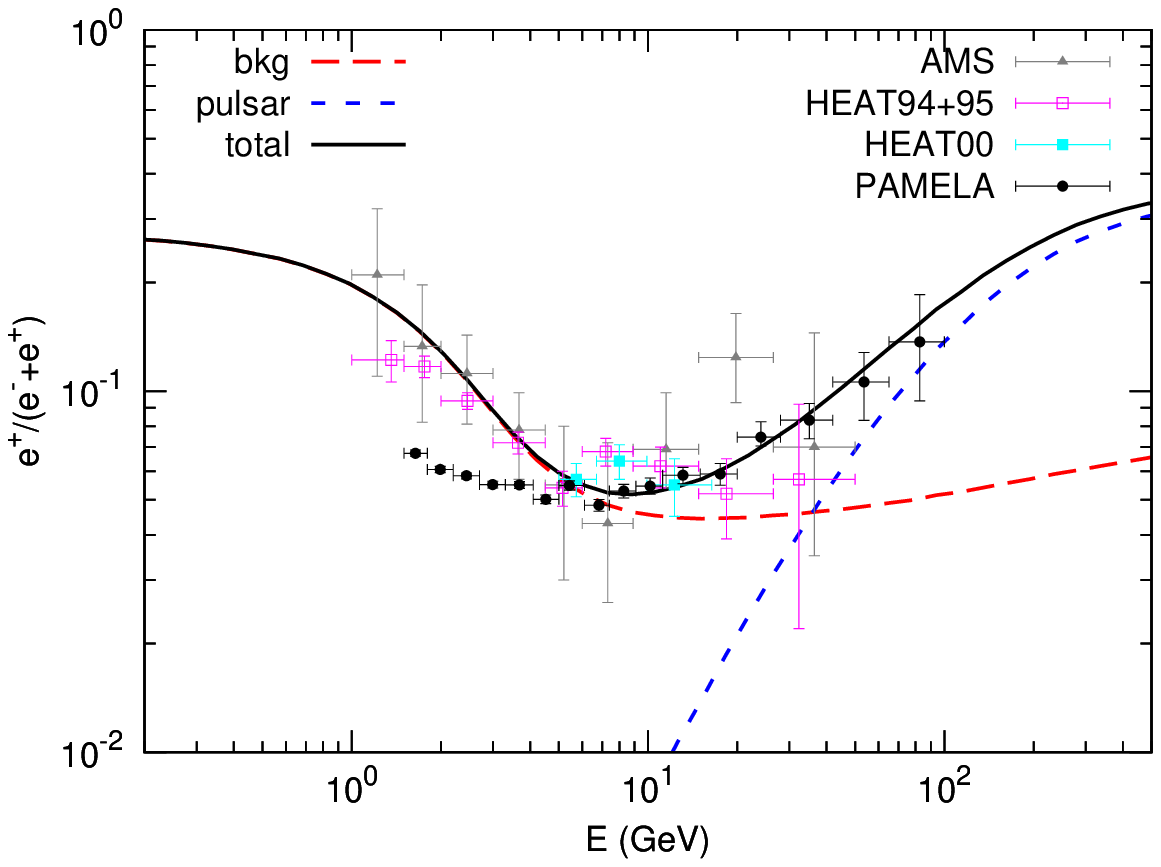}
\includegraphics[width=0.45\columnwidth]{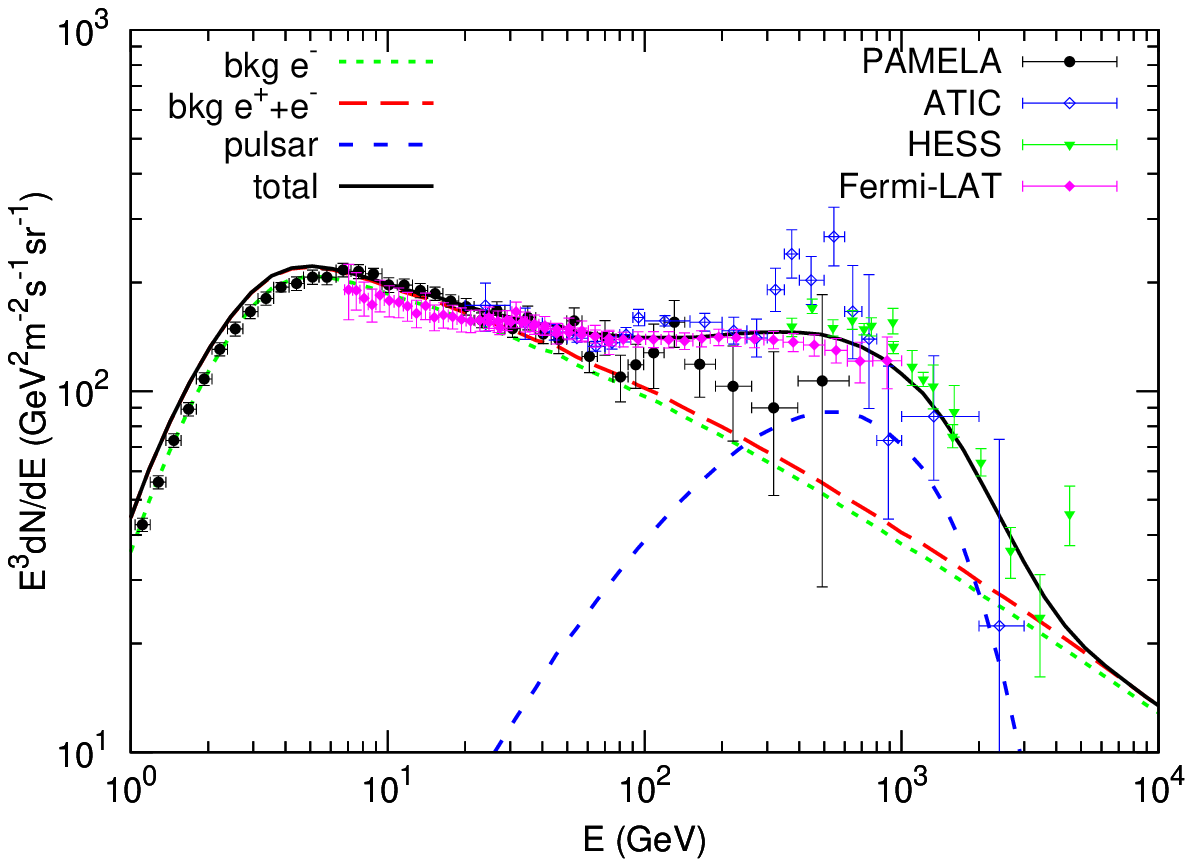}
\caption{Same as Fig. \ref{fig:epm_psr} but for the model with pulsars as
the extra sources of $e^+e^-$.
\label{fig:epm_psr}}
\end{center}
\end{figure}

%The 1-dimensional probability distributions of the nine parameters are
%shown in Fig. \ref{fig:pulsar1d}. 
Compared with the conventional GALPROP
background model, $\gamma_1=1.63\pm0.08$ is consistent with $1.60$
as given in \cite{2004ApJ...613..962S}. Note that there is a
correlation between parameters $\gamma_1$ and the modulation potential
$\phi$, as shown in Fig. \ref{fig:pulsar2d}.
The high energy injection spectrum $\gamma_2=2.74\pm0.02$ is softer than
$2.54$ as adopted in the conventional model \cite{2004ApJ...613..962S}.
In the conventional model only the background contribution is employed
to fit the data, while here an additional component from the extra sources
are added together. Thus the high energy spectrum can be much softer.
This might be important for the understanding of the background
contribution to the CR electrons.

There is a factor $c_{e^+}\approx 1.6$ needed for the background
positrons to fit the data. As discussed above, such a factor may be
ascribed to the uncertainties of the propagation parameters, the ISM
density and the strong interaction cross sections. Those uncertainties
may be energy-dependent and can not be simply described with a constant
factor (see e.g., \cite{2009A&A...501..821D}). For example in this
work we use the parameterization given in \cite{2006ApJ...647..692K}
(Kamae06) to calculate the $pp$ interaction to generate positrons.
Compared with the $pp$ collision model Badhwar77 \cite{1977PhRvD..15..820B}
as adopted in GALPROP, the Kamae06 model gives systematically fewer
positrons, especially for energies from several to tens of GeV
\cite{2009A&A...501..821D}. If these input uncertainties are better
understood in the future, we can include the background positron
production with a more general form in the MCMC fit.

The fitted parameters for pulsars are $\alpha=1.201^{+0.120}_{-0.129}$,
$E_c=0.797^{^+0.154}_{-0.164}$ TeV. The constraints of the pulsar
parameters are weaker when compared with the background parameters.
This is because the cutoff energy is mainly determined by HESS data.
However, for HESS data, the very large systematic errors
due to poor absolute energy calibration (not shown in Fig.
\ref{fig:epm_psr}) makes it impossible to precisely determine $E_c$.
In the calculation the relative systematic uncertainty of the flux is
estimated to be $(\Gamma-1)\Delta E/E$ \cite{2010PhRvD..82i2004A}.
For $\sim 15\%$ uncertainty of the energy scale, the flux uncertainty
is $\sim 30\%$ and $\sim 45\%$ for energies below and above $\sim 1$ TeV.
Due to the correlations between $E_c$ and $\alpha$, $A_{\rm psr}$,
the other two parameters are also relatively less constrained.

\subsection{Dark matter annihilation scenario}

First we consider the case without including the antiproton data.
%The one dimensional probability distributions of the model parameters
%are shown in Fig. \ref{fig:dmnopbar1d}. It is shown 
We found that the background
parameters are similar with the pulsar scenario. The overall $\chi^2$
value of the best-fit is about $90.2$, which is very close to
$\chi^2_{\rm psr}\approx 91.6$. That is to say both the astrophysical
scenario and DM scenario can give comparable fit to the $e^+e^-$ data.
We may not be able to discriminate these two scenarios from the $e^+e^-$
spectra \cite{2009PhLB..678..283B,2009PhRvD..80f3005M,2009arXiv0909.1182C,
2009PhRvD..79d1301P,2010JCAP...12..020P}, but need to resort to other
probes such as the $e^+e^-$ anisotropy \cite{2010APh....34...59C}
%, charge asymmetry \cite{Rojas:2011wx}
and $\gamma$-rays \cite{2009PhRvD..80b3007Z}.

%\begin{figure}[!htb]
%\begin{center}
%\includegraphics[width=0.9\columnwidth]{dmnopbar1d.ps}
%\caption{Same as Fig. \ref{fig:pulsar1d} but for the DM annihilation
%scenario. The antiproton data are not included in the MCMC fit.
%\label{fig:dmnopbar1d}}
%\end{center}
%\end{figure}

The fitted mass and cross section of DM are $m_{\chi}=2.37^{+0.47}_{-0.39}$
TeV, $\langle\sigma v\rangle=4.7^{+1.6}_{-1.2}\times 10^{-23}$ cm$^3$
s$^{-1}$. Similar as the pulsar case, these parameters are also less
constrained due to the large uncertainties of the HESS data. We can
also determine the branching ratios to different flavors of leptons:
$B_e(2\sigma)<0.37$, $B_{\mu}(2\sigma)<0.38$ and $B_{\tau}=0.69^{+0.14}
_{-0.15}$. These results are qualitatively consistent with our previous
fit in \cite{2010PhRvD..81b3516L}, but with larger uncertainties because
we include the systematic errors of HESS data in the present work.
We should be cautious that the determination of the branching ratios
may suffer from systematic uncertainties of the background electrons
and positrons. As discussed in the previous sub-section, the present
understanding of the background positrons is far from precise. Therefore
the quoted fitting uncertainties of the branching ratios may be
under-estimated.

%\begin{figure}[!htb]
%\begin{center}
%\includegraphics[width=0.9\columnwidth]{dmpbar1d.ps}
%\caption{Same as Fig. \ref{fig:dmnopbar1d} but with antiproton data.
%\label{fig:dmpbar1d}}
%\end{center}
%\end{figure}

\begin{figure}[!htb]
\begin{center}
\includegraphics[width=0.9\columnwidth]{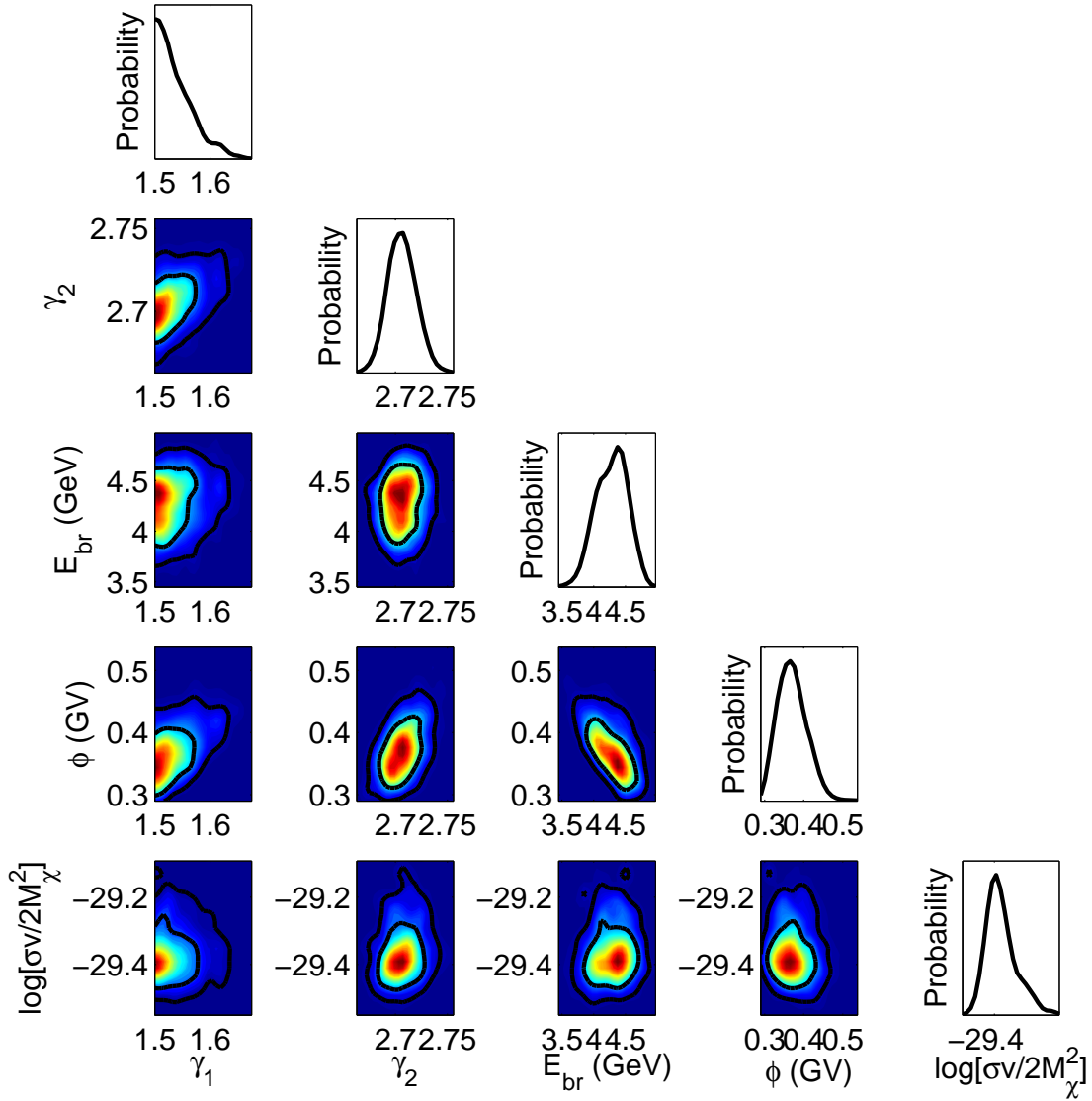}
\caption{Same as Fig. \ref{fig:pulsar2d} but for the DM scenario with
antiproton data.
\label{fig:dm2d}}
\end{center}
\end{figure}

\begin{figure}[!htb]
\begin{center}
\includegraphics[width=0.6\columnwidth]{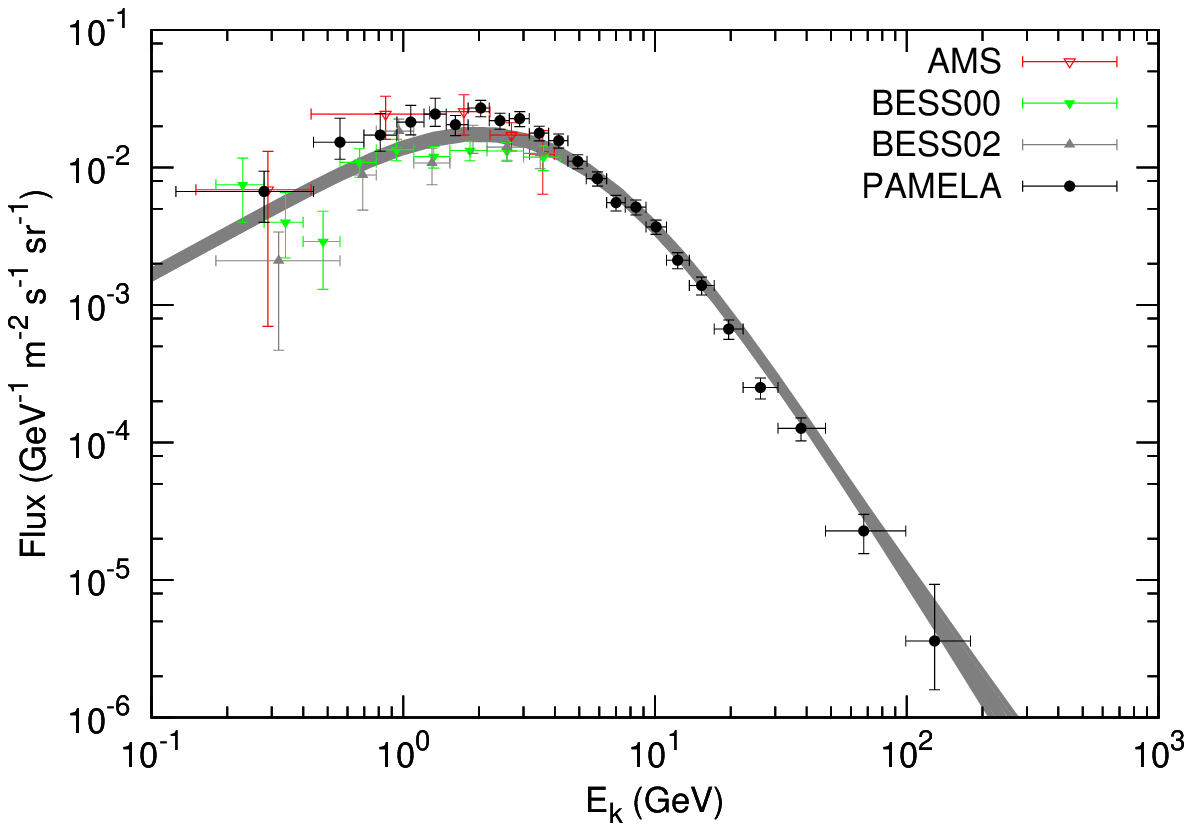}
\caption{Fitted $2\sigma$ range of the antiproton fluxes of the DM
annihilation scenario. References of the data are:
AMS \cite{2002PhR...366..331A}, BESS00 \cite{2002PhRvL..88e1101A},
BESS02 \cite{2005ICRC....3...13H}, PAMELA \cite{2010PhRvL.105l1101A}.
\label{fig:antip_dm}}
\end{center}
\end{figure}

Then we consider the case with antiproton data. The coupling to quarks 
of DM annihilation is taken into account. The two dimensional constraints
of part of parameters are present in Fig.
\ref{fig:dm2d}. Compared with the case without antiproton data, we
find that the background parameters are slightly different. The reason 
for the change is mainly due to the low energy spectrum of the background 
antiprotons. As shown in Fig. \ref{fig:antip_dm} the calculated secondary 
(including tertiary) antiprotons are basically consistent with the PAMELA 
data for energies higher than several GeV, but under-estimate the flux 
for the low energy part. This problem was pointed out years ago
\cite{2002ApJ...565..280M,2003ApJ...586.1050M}, that in the diffusive 
reacceleration model which gives the best fit to the B/C data the 
antiprotons are under-estimated. 
There seems to be a contradiction between the B/C and the antiproton
data\footnote{Note, however, in \cite{2001ApJ...563..172D} the calculated 
antiproton flux based on the propagation parameters fitted according to the 
B/C data \cite{2001ApJ...555..585M} was consistent with the observational 
data. As pointed out in \cite{2003ApJ...586.1050M} the fit in 
\cite{2001ApJ...555..585M} was actually based on the high energy data. 
And in their results the low energy B/C was in fact over estimated. 
Furthermore the antiproton production cross sections would also lead to
uncertainties \cite{2003ApJ...586.1050M}. Alternatively, in
\cite{2010APh....34..274D} a unified model to explain the B/C and
antiproton data was proposed, with an empirical modification of the 
diffusion coefficient at low speed of particles.}. To better understand 
this issue we may need more precise measurement about
the B/C data. Back to this work, a lower antiproton flux at low
energies will require a smaller solar modulation potential, and
therefore the background parameters $\gamma_1$, $\gamma_2$ and
$E_{\rm br}$ will change accordingly due to the correlations among
them as shown in Fig. \ref{fig:dm2d}.

One thing important is that we can derive a self-consistent constraint
on the quark branching ratio of DM annihilation. The marginalized
$2\sigma$ upper limit of $B_u$ is about $0.5\%$. Compared with our
previous study \cite{2009arXiv0911.1002L} the upper limit of $B_u$ is
several times smaller. This is probably due to the bad fit to the
antiproton data with the background model. We have run a test to employ
the background antiproton spectrum used in \cite{2009arXiv0911.1002L}
and found that the $2\sigma$ upper limit of $B_u$ is about $2.1\%$,
which is consistent with the previous results. This means that the
current constraint on $B_u$ is systematics dominated. Better understanding
of the background contribution to the antiproton flux is necessary to
further address this issue. We plot the $2\sigma$ range of the antiproton
fluxes in Fig. \ref{fig:antip_dm}.

\section{Conclusion and Discussion}
\label{summary}

Recently more and more observational data of CRs with unprecedented
precision are available, which makes it possible to better approach
the understanding of the basic problems of CRs. Based on MCMC analysis
in our previous work \cite{2010PhRvD..81b3516L,2009arXiv0911.1002L},
we embed GALPROP and PYTHIA into MCMC sampler and study the implication
of the newest CR data, including the positron fraction, electrons
(pure $e^-$ and $e^+e^-$) and antiprotons from PAMELA, Fermi-LAT
and HESS experiments in this work.
%With this {\it CosRayMC} package, we can do further research on the
%constrains of propagation parameters.

We work in the frame of diffusive reacceleration propagation model of
CRs. The propagation parameters are adopted according to the fit to
currently available B/C data \cite{2011ApJ...729..106T}, with a slight
adjustment of the Helium abundance to better match the PAMELA data
\cite{2011Sci...332...69A}. We find that the pure background to explain
the CR $e^+e^-$ data is disfavored with a very high significance.
Therefore it is strongly implied that we may need some extra sources
to produce the positrons/electrons.

We then consider two different scenarios, which are widely discussed 
in recent literature, to explain the $e^+e^-$ excesses. One is the 
astrophysical scenario with pulsars as the typical example, and the 
other is the DM annihilation scenario. Using the global fitting method, 
we can determine the parameters of both the background and the extra 
sources. The low energy spectral index of the background electrons is 
consistent with the conventional model adopted in the previous study, 
while the high energy index is softer in this work. We find that, 
together with the background contribution, both of these scenarios can 
give very good fit to the data. For the case with only the $e^+e^-$ data, the
goodness-of-fit of the pulsar scenario and DM scenario are almost
identical. With the antiproton data included, we can further constrain 
the coupling to quarks of the DM model. The branching ratio to quarks 
of the DM annihilation final states is constrained to be $<0.5\%$ at 
$2\sigma$ confidence level.

\begin{figure}[!htb]
\begin{center}
\includegraphics[width=0.6\columnwidth]{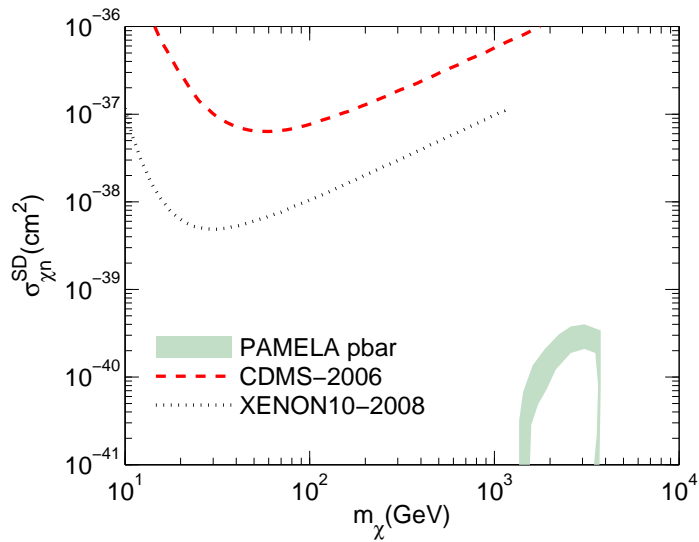}
\caption{Constraints on the spin-dependent DM-neutron scattering
cross section from the PAMELA antiproton data assuming axial
vector interaction form of DM particles and SM fermions. The
shaded region represents the uncertainties of the astrophysical
background of antiprotons. Also shown are the results from CDMS
\cite{2006PhRvD..73a1102A} and XENON10 \cite{2008PhRvL.101i1301A}.
\label{fig:direct_comp}}
\end{center}
\end{figure}

The constraint on the quark branching ratio (or the cross section to quarks) 
allows us to make a comparison with the direct detection experiments, 
assuming some effective interaction operators between DM and standard model 
(SM) particles. As shown in \cite{2010arXiv1012.2022Z}, generally the 
constraint of DM-SM coupling from direct experiments is much stronger 
than the indirect search for spin-independent interactions. However, for
the spin-dependent interactions, the indirect search constraint
could be comparable or stronger than the direct experiments. In Fig. 
\ref{fig:direct_comp} we give the constraints on the spin-dependent 
DM-neutron scattering cross section according to the cross section 
of DM annihilation to quarks, assuming the axial vector interaction 
form of DM particles and SM fermions. The ratio between 
$\langle\sigma v\rangle$ and $\sigma_{\chi n}^{\rm SD}$ is $5c
(1+m_n/m_{\chi})^2/m_n^2\sum_{q}\sqrt{1-m_q^2/m_{\chi}^2}m_q^2$
\cite{2010arXiv1012.2022Z}, where $c$ is light speed, $m_n$ is the
mass of neutron and $m_q$ is the mass of quark. The sum is for all
the flavors of quarks. Considering the uncertainties of the background 
antiprotons, we show the results of two antiproton background models 
with shaded region: the result calculated in this work and the one 
used in \cite{2009arXiv0911.1002L}. The results show that the 
constraint on the spin-dependent cross section between DM and nucleon 
from PAMELA antiproton data is much stronger than the direct experiments.

Although the existence of the extra sources is evident, we still can
not discriminate the pulsar and DM scenarios right now. These two models
both can fit the data very well, with $\chi^2_{\rm red}=0.833$ and
$\chi^2_{\rm red}=0.827$ respectively. Other new probes such as
the $e^+e^-$ anisotropy and $\gamma$-rays are needed to distinguish
these models.

The $\gamma$-rays should give further constraints on the models,
especially for the DM annihilation scenario. However, the diffuse
$\gamma$-ray data of Fermi-LAT are still in processing, and the
modeling of the $\gamma$-ray data seems not trivial. Furthermore
the $\gamma$-rays are more sensitive to the density profile of DM,
which is not the most relevant parameter of the present study.
Therefore we do not include the $\gamma$-ray data. It should be a
future direction to include the $\gamma$-ray data in the
{\it CosRayMC} package.

There are some systematical uncertainties of the current study, which are
mainly due to the lack of knowledge about the related issues and can be
improved in the future. First, the solar modulation which affects the low
energy spectra of CR particles ($E\lesssim 30$ GeV) if not clear. In this
work we use the simple force-field approximation \cite{1968ApJ...154.1011G}
to deal with the solar modulation. But such a model seems to fail to explain
the low energy data of the PAMELA positron fraction, which may indicate
the charge dependent modulation effect \cite{1996ApJ...464..507C,
2009NJPh...11j5021B,2010arXiv1011.4843B}. Second, the propagation model
is adopted as the diffusive reacceleration model with the parameters
best fitting the B/C data \cite{2011ApJ...729..106T}. There are
uncertainties of the propagation parameters. Furthermore the diffusive
reacceleration model may also have some systematical errors when comparing
with all of the CR data. For example the antiprotons below several GeV
is under-estimated in this model. We need to understand the propagation
model better with more precise data. Third, the secondary positron and
antiproton production may suffer from the uncertainties from the ISM
distribution and the hadronic cross sections. All of these uncertainties
may affect the quantitative results of the current study. Nevertheless,
with more and more precise data available in the near future (e.g., AMS02,
which was launched recently), and better control of the above mentioned
systematical errors, it is a right direction to do the global fit to
derive the constraints and implication on the models.

\section*{Acknowledgements}
We thank Hong-Bo Hu and Zhao-Huan Yu for helpful discussion.
The calculation is taken on Deepcomp7000 of Supercomputing Center,
Computer Network Information Center of Chinese Academy of Sciences.
This work is supported in part by National Natural Science Foundation
of China under Grant Nos. 90303004, 10533010, 10675136, 10803001, 
11033005  and 11075169, by 973 Program under Grant No. 2010CB83300,
 by the Chinese Academy of Science under Grant No. 
KJCX2-EW-W01 and by the Youth Foundation of the Institute of High Energy
Physics under Grant No. H95461N.

\bibliographystyle{apsrev}
\bibliography{refs}

\end{document}